\documentclass[a4paper, 10pt, conference]{ieeeconf}      

\IEEEoverridecommandlockouts                              

\usepackage{graphicx}									  
\usepackage{siunitx}
\usepackage{amsmath}
\usepackage{hyperref}

\usepackage{xcolor}

\usepackage{algorithm}
\usepackage[noend]{algpseudocode}

\title{\LARGE \bf
V2V communication-based rail collision avoidance system \\for urban light rail vehicles*
}

\author{Loi Do$^{1}$, Ivo Herman$^{2}$ and Zden\v{e}k Hur\'ak$^{3}$
\thanks{*This work was funded by Technology Agency of the Czech Republic within the program Epsilon, the project TH03010155.}
\thanks{$^{1}$Loi Do is a doctoral student at Faculty of Electrical Engineering, Czech Technical University in Prague.
        {\tt\small doloi@fel.cvut.cz}}%
\thanks{$^{2}$Ivo Herman is with Herman Systems, Ltd.
        {\tt\small ivo.herman@herman.cz}}%
\thanks{$^{3}$Zden\v{e}k Hur\'ak is with Faculty of Electrical Engineering, Czech Technical University in Prague.
        {\tt\small hurak@fel.cvut.cz}}%
}

\begin{document}
\maketitle
\thispagestyle{empty}
\pagestyle{empty}

\begin{abstract}
In this paper, we document a design, implementation, and field tests of a vehicle-to-vehicle (V2V) communication-enabled rail collision avoidance system (RCAS) for urban light rail vehicles---trams. 
The RCAS runs onboard a tram and issues an acoustic warning to a tram driver if a collision with another tram is imminent---no commands to the braking subsystem are issued in the current version.
The prediction of an imminent collision with another tram is based on real-time evaluation of predicted trajectories of both trams. The predictions are based on mathematical models of the longitudinal dynamics of the vehicles and real-time estimation of the current motion states (position and velocity).
The information about the other tram's predicted trajectory is accessed through V2V communication.
We also document the results of verification of the functionality of the proposed RCAS through several field tests with a real tram. 
\end{abstract}


\section{Introduction}
\subsection{Goal and Motivation}
In this paper, we present a collision avoidance/warning system for trams---light rail vehicles designed for passenger transportation, mainly in urban areas. 
Unlike heavy rail vehicles, tram tracks run along urban streets, thus frequently interact with other road vehicles or pedestrians.
In addition, in cities with a dense tram network, both the distance and the time gaps between two trams go down to a few meters and seconds (e.g., at tram stops), respectively.
These conditions result in a high number of collisions.
Every year, Prague Public Transit, Co. Inc. registers in total well above $\num{1000}$ of collisions, of which around $\num{350}$ collisions are tram-to-tram collisions (the last accident of the latter type happened two weeks before the conference deadline). 
This motivates the development of various collision avoidance systems; if tram-to-tram collisions are considered only, the technology is referred to as \textit{rail collision avoidance system} (RCAS). The paper documents one such development.

Restriction of the focus to collisions between trams offers major advantages for the development. Both participants in the (potential) collision are operated by a single company, which makes coordinated collision avoidance schemes (almost) directly realizable.
The technology enabling such coordination is WLAN-based (IEEE 802.11p) vehicle-to-vehicle (V2V) communication, which typically comes as one mode of a broader vehicle-to-everything (V2X) communication. This technology is often referred to as dedicated short-range communication (DSRC). Through this communication, trams can exchange their position and velocity estimates and future trajectory predictions. Since possible collisions with other trams can be predicted onboard every tram, the resulting RCAS can be regarded as fully distributed.

This particular project was realized in close cooperation with a public transportation operator in Ostrava (Dopravni podnik Ostrava, DPO), Czech Republic. The company has equipped all their vehicles with V2X on-board units (OBUs). 
Their initial motivation for equipping their fleet with V2X on-board units was to: enable/upgrade public transportation priority schemes (a tram or a bus approaching an intersection can request a priority from the traffic light controller), to automate some traffic-related tasks (opening the gates, for example) and general information sharing between vehicles. 
With the V2X hardware and software already deployed, RCAS based on V2X communication follows naturally as a cost-effective way to increase the safety of the tram traffic. Although a full RCAS would require not only the communication between the vehicles but also some direct measurement of a distance between the vehicles, in this project, we investigated the possibility to base such RCAS without the need to install any additional hardware (a radar, a lidar or a camera). 

The presented system could better be named \textit{rail collision warning system} because currently, no commands to the braking subsystem are issued. The system warns the driver acoustically if a collision with another tram is imminent.


\subsection{State of the Art}
The concept of collision avoidance (warning) systems (CAS) for vehicles based on V2V communication is well documented in several works.
As the trams combine the aspects of both road and rail transportation, we present a short description of state of the art from both transportation domains.
Regarding the road vehicles, it has been shown that communication between the vehicles introduces several advantages over vision-(only)-based (radar, lidar, etc.) approaches~\cite{xiang_research_2014}.

For many types of rail transportation systems (heavy-rail, metro), CAS are usually based on central monitoring~\cite{wang_parallel_2015}, although the use of V2V communication-based RCAS was also reported~\cite{strang_pdf_2006},~\cite{lehner_reliable_2009}.
The use of V2V communication in collision avoidance between rail and road vehicles at railroad crossing was analyzed in~\cite{choi_measurements_2018}.
Not many references can be found explicitly for trams---in fact, we could only find a report on the use of laser-scanner~\cite{katz_towards_2013}.

\section{Existing system in Ostrava}
Our exploration of the problem of tram-to-tram collision avoidance was since the very beginning motivated by the existing fleet and its V2X infrastructure. Therefore, it may be useful to share some details of what could be truly regarded as a large-scale V2X testbed. 
The municipal public transportation company of Ostrava acquired recently (in 2019) a V2X (vehicle-to-everything) solution for its whole fleet, consisting of more than 600 vehicles. 
The fleet comprises 250 trams, 300 buses, and 70 trolleybuses. 
The challenge was that the fleet is of various age, especially the age of trams ranges from 1960 (T3 trams) to modern Stadler trams. In addition, there are more than 30 types of vehicles. 
However, all these vehicles are equipped with V2X units, enabling the realization of large-scale experiments.

The onboard units (OBU) were developed by Herman Systems, Ltd., and fully conforms to all ETSI norms and C-ROADS standards. 
In addition, the OBUs are closely integrated to other onboard systems, mainly the board computer. 
Such integration allows, except for the scenario described below, a realization of many interesting use-cases, increasing both the safety and attractiveness of public transportation. The already working use-cases are:
\begin{itemize}
    \item Public transportation priority -- the vehicle requests public transportation priority at intersections both using older proprietary radio systems and the new V2X system via SRM messages. The intersections use either the old or the new system. 
    The intersection controller actively cooperates via V2X with the vehicle at a stop to guarantee that after it leaves the stop, the vehicle will pass on a green light.  
    \item Line connections -- the vehicle periodically broadcasts its public transportation status (current line, connection, and destination) in a CAM message. Then other vehicles "know" which vehicles are nearby, so the driver can better decide whether to wait for a connecting line or not.
    \item Warning of individual traffic drivers -- in potentially dangerous situations, the public transportation vehicle transmits a DENM message. This is used at dangerous public transportation stops (for instance, when the trams stop at the middle of the roadway), 
    at dangerous crossings of railway and roadway (typically places, where tram enters/leaves a roadway) or when the driver leaves the vehicle to switch the rail switch manually.     
\end{itemize} 
To increase the safety of public transportation itself, the public transportation company also allowed us to test the possibility of V2X-based collision avoidance, which is described in next.

\subsection{Onboard units}
The installed onboard units are of type UCU 5.0V. 
The units are equipped with a powerful V2X module, a mobile broadband connection, and of course, a GNSS module. 
The unit also has an integrated inertial measurement unit (IMU).
What makes the solution in DPO unique is the ability of OBUs to obtain the true vehicle state, such as odometric speed or activation of brakes, from all types of vehicles. 
For new vehicles, this is quite easy, a CAN bus is usually available, and for buses, it is even standardized (norm SAE J1939). 
But as we mentioned above, the fleet also consists of very old trams, which have no internal bus.
To cope with that, the OBU is connected to the tachograph for trams. 
The tachograph provides not only odometric speed but also other states of the vehicle - activation of brakes, turning indication, driving direction, etc. 
Connection to tachographs allows data acquisition for different vehicle types in a unified way.

To enable good tracking of a vehicle and also a generation of warning (DENM) messages, the unit also provides a localization on a digital map. 
From the standard ETSI messages, the unit supports the generation of CAM, DENM, SRMs (for priority at intersections), and for receptions and presentation to the driver IVI, MAP (Message with detailed road topology information), SPAT (Signal Phase And Timing) and SSM. 
The unit and the V2X stack it runs passed ETSI Plugtest. All units are also ready for the integration of security and PKI communication. 

For the purpose of collision avoidance system, the relevant properties of UCU 5.0V are mainly the ability to measure the quantities describing the motion of a vehicle:
\begin{itemize}
	\item Absolute geographic position in WGS84 coordinates.
	\item Longitudinal speed (from a GNSS module or a tachograph).
	\item Longitudinal acceleration measured by IMU.	
\end{itemize}
All measurements except the speed from tachograph are sampled with $\SI{10}{\hertz}$, while the odometric speed from tachograph are sampled with frequency $\num{2}-\SI{3}{\hertz}$, depending on the type of a tachograph (in the experiment we used $\SI{2}{\hertz}$).

\subsection{DSRC messages and properties}
We believe that DSRC communication technology is quite known in the community, so the description here will be focused on the properties required for the goal of this paper. 
We refer the reader to \cite{festag_cooperative_2014} for a thorough description.
The DSRC (V2X) technology in Europe is based on the standard IEEE 802.11 and operates in $\SI{5.9}{\giga\hertz}$ band (formerly known as 802.11p). 
The European set of standards is called ETSI ITS-G5. 
The standards comprise all the communication layers from the access layer to the application (called facility) layer. 
The network layer is called GeoNetworking and allows geographic-location-based routing, where surrounding V2X units act as routers. 
From the facility layer, two messages can be used for collision-avoidance:
\begin{itemize}
    \item CAM message (Cooperative Awareness Messages)
    \item DENM message (Decentralized Environmental Notification message)    
\end{itemize}
Transmission of CAM messages is a way with which vehicles exchange information about their state (position, velocity, heading, and also public transportation state = line, connection, ...). 
These messages are transmitted at least once a second, but if the vehicle state changes (change of a position, speed, or heading), CAMs shall be generated with a higher frequency with a maximum of $\SI{10}{\hertz}$. 
In our system, we use CAM messages to detect that a tram is ahead.

DENM messages are then generated when an imminent collision is detected. A message of Longitudinal Collision Warning is sent in order to warn other vehicles that abrupt braking might occur. 
This message shall then be displayed on a board computer display.

\section{Method Description}
\begin{figure}[tb]
	\centering
	\includegraphics[width=8.4cm]{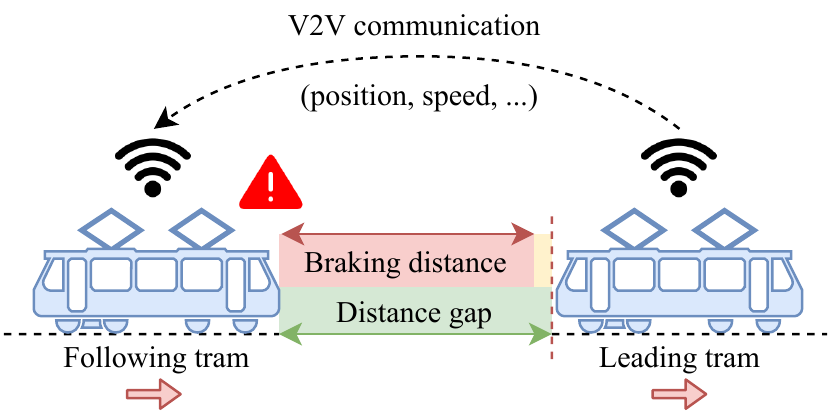}
	\caption{Scheme of the proposed RCAS for a case of an imminent rear-end collision.}\label{fig:algorithm_visual}
\end{figure}

In general, there are three types of tram-to-tram collisions: rear-end, flank, and head-on.
We propose a RCAS which runs onboard a tram and can avoid such collisions by predicting the trajectories of both trams and warns the driver promptly in case of an imminent collision.
From the perspective of a single tram, the collision avoidance scheme consists of the following tasks (see the Fig.~\ref{fig:algorithm_visual} for the scheme of the algorithm):
\begin{itemize}
\item Onboard of a (following) tram runs an estimator of its absolute (geographic) position and longitudinal (along a tram track) speed. 
\item The estimated and measured quantities (position, speed, and heading) and length of the tram are then broadcast via V2V communication while the same quantities are received from the leading tram.
\item OBU then predicts the braking trajectory of the following tram and the trajectory of the leading tram (assuming constant speed) using the received information.
\item If the collision of predicted braking trajectory and the trajectory of the following tram is detected, RCAS warns the driver to promptly start braking to avoid a collision.
\end{itemize}
We describe three main tasks of RCAS: prediction of a tram's longitudinal motion, braking distance prediction, and finally, the collision prediction in detail below.


\subsection{Position and Speed Estimation}
We estimate the absolute position and the longitudinal speed (along the track) using linear (discrete) Kalman filter (KF) augmented with the use of the digital map~\cite{liu_generating_2013}.
The set of equations giving recursive estimate of a state $x_{k}$ at time $k$ is:
\begin{subequations}\label{eq:kalman_filter}
\begin{align}
x_{k|k-1} 	&= F_{k-1} x_{k-1} + G_{k-1} u_{k-1}\;, 		\\
P_{k|k-1} 	&= F_{k-1} P_{k-1} F_{k-1}^T + Q_{k-1}\;,		\\
K_{k} 		&= P_{k|k-1}H_k^T(H_k P_{k|k-1}H_k^T + R_k)^{-1}\;,	\\
x_k 		&= x_{k|k-1} + K_k(y_k - H_k x_{k|k-1})\;,		\\
P_k			&= P_{k|k-1} - K_k H_k P_{k|k-1}\;.
\end{align}
\end{subequations}
Matrices $F_k$, $G_k$, $Q_k$, $H_k$, $R_k$, are parameters of the model given in~\eqref{eq:state_space_matrices} and \eqref{eq:cov_matrices} , $y_k$ is a vector of measurements and $P_k$ is error covariance matrix.
For a detailed description of the general concept of the KF, we refer the reader to one of many publications, for instance~\cite{kailath_linear_2000}.

\begin{figure}[b]
	\centering
	\includegraphics[width=8.4cm]{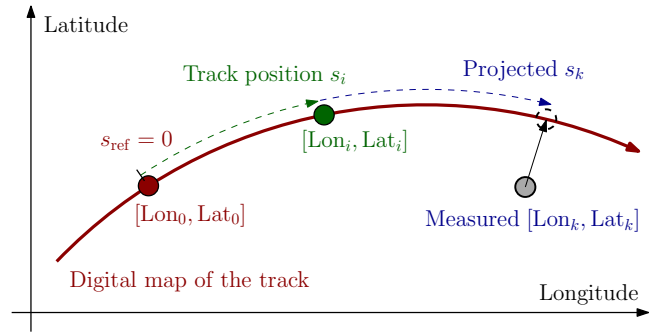}
	\caption{An illustration of the use of the digital map for transformation between track position $s_i$ and absolute position $[\mathrm{Lon}_i, \mathrm{Lat}_i]$.}\label{fig:GPS_projection_onto_track}
\end{figure}

The use of the digital map in the estimation scheme is depicted in Fig.~\ref{fig:GPS_projection_onto_track}.
Instead of direct estimation of the two-dimensional position $i$ of the tram given by coordinates $[\mathrm{Lon}_i, \mathrm{Lat}_i]$, we estimate the one-dimensional track position $s_i\,[\si{\meter}]$ (i.e. total traveled distance from reference point on the track).
Then, the track position $s_i$ can be easily transformed into an absolute position as the digital map (with $[\mathrm{Lon}_0, \mathrm{Lat}_0]$ as fixed reference point) provides one-to-one mapping from $s_i$ to $[\mathrm{Lon}_i, \mathrm{Lat}_i]$.
Analogously, the measured absolute position $[\mathrm{Lon}_k, \mathrm{Lat}_k]$ can be represented by a track position $s_k$ by computing the track distance from reference point to the orthogonal projection of a measured point onto the track.
The reference point $[\mathrm{Lon}_0, \mathrm{Lat}_0]$ is initialized at the start of the estimation with first (relevant) GNSS position measurement.
The use of the digital map, however, imposes a problem of ambiguous transformation of the track position $s$ into the absolute position at track switches where the track divides into several branches.
We address this problem by selecting one of the branches which correspond the most to the latest GNSS measurements. 
When the estimated position with selected branch does not correspondent (in next several iterations) to a position measurement (measured position is closer to another branch), the KF is reset.

It is thus sufficient to estimate only the track position $s$. 
We define $x_{k} = \left[s_k, v_k, a_k\right]^T$ as vector of track position, longitudinal velocity and acceleration, respectively.
As state-space model used in~\eqref{eq:kalman_filter} for the estimation of the state, considering the unknown input (OBU does not have access to the value of throttle notch nor the electrical current in motors), we use the \textit{constant acceleration model} defined by matrices:
\begin{equation}\label{eq:state_space_matrices}
F_k =
\begin{bmatrix}
1	&	\Delta t	&	0.5\Delta t^2	\\
0	&	1			&	\Delta t	\\
0	&	0			&	1	\\
\end{bmatrix}
\;,
G_k = 0\;, 
H_k =
I_{3\times3}\;,
\end{equation}
where $I_{3\times3}$ is $3\times3$ identity matrix and $\Delta t = \SI{0.1}{\second}$ is a sample time (corresponding to the sample time of the GNSS module in the OBU).
Covariance matrices of the model are:
\begin{equation}\label{eq:cov_matrices}
Q_k = q
\begin{bmatrix}
\Delta t^{5}/20     & \Delta t^{4}/8  & \Delta t^3/6	\\
\Delta t^{4}/8      & \Delta t^{3}/3  & \Delta t^2/2	\\
\Delta t^{3}/6      & \Delta t^{2}/2  & \Delta t	\\
\end{bmatrix},
R_k =
\begin{bmatrix}
\sigma_1     & 0  & 0	\\
0      & \sigma_2  & 0	\\
0      & 0 & \sigma_3	\\
\end{bmatrix}
\end{equation}
where $q$ and $\sigma_i$ are parameters to be defined.

Note that in general, not all measurements are available at time $k$ (temporary lost of GNSS signal, tachograph sampled with $\SI{2}{\hertz}$).
Thus, the matrices $H_k$ and $R_k$ are time-dependent.
For instance, if the position measurement is not available at time $k$, the matrices $H_k$ and $R_k$ become:
\begin{equation}
H_k = 
\begin{bmatrix}
0 & 1 & 0 \\
0 & 0 & 1
\end{bmatrix}\;,
R_k =
\begin{bmatrix}
\sigma_2  & 0	\\
0 & \sigma_3	\\
\end{bmatrix}\;.
\end{equation}

\subsection{Braking Trajectory Prediction}\label{sec:braking_traj_pred}
To predict the braking trajectory of a tram, we use the model-based method described in~\cite{do_onboard_2020}. 
As shown, this method better estimates the total braking distance (and thus also the whole trajectory) in the presence of varying weight or adhesion conditions.
The method, including deviations caused by different tram type, is briefly summarized below.

The system of equations describing the model of dynamics of a tram is~\cite{do_onboard_2020}:
\begin{subequations}\label{eq:braking_dist_model}
\begin{align}
\tilde{T}_\mathrm{mot} &= 
\begin{cases}
K_tp								&\text{for }T_\mathrm{mot}\omega_\mathrm{wh}	<  		P_\mathrm{max} \;,\\
P_\mathrm{max}/\omega_\mathrm{wh}	&\text{for }T_\mathrm{mot}\omega_\mathrm{wh}	\geq  	P_\mathrm{max} \;,\\
\end{cases}
\\
\dot{T}_\mathrm{mot}(t) &= 3\left(\tilde{T}_\mathrm{mot} - T_\mathrm{mot}\right)\;,
\\
\dot{\omega}_\mathrm{wh}(t) &= 2m_w^{-1}r^{-2} \left(T_\mathrm{mot}(t) - r\mu(v_s) Mg \right)\;, 
\\
M\dot{v_t}(t) &= \mu(v_s) Mg - F_\mathrm{r} - Mg\sin\theta \;, 
\\
F_\mathrm{r} &= 0.0147M + 125.83v_t(t)\;.
\\
\mu(v_s) &= c_a e^{-a_av_s} - d_ae^{-b_av_s}\;,
\\
v_s &= r\omega_\text{wh}(t) - v_t(t) \;.
\end{align}
\end{subequations}
The values of $M$, $m_w$, $r$, $K_t$  and $P_\mathrm{max}$ (tram weight, wheel mass, and radius, traction constant and maximal power of the motors, respectively) depend on the tram and need to be identified to predict the braking trajectory correctly. 
The values of $a_a, b_a, c_a$ and $d_a$ describe the adhesion conditions.

For the prediction of braking distance at time $t_0$, we set currently estimated speed as the initial speed  $v_t(t_0)$ and notch for maximal service braking, that is $p=-7$.
The braking trajectory is then defined by the set of points (geographic coordinates) in time from the start of the simulation until the longitudinal speed of tram is zero, $v_t(t)=0$. 

\subsection{Collision prediction}
With estimated position and speed onboard of both trams, it is now straightforward to detect an imminent collision.
From the perspective on one tram, at every time step, the OBU estimates tram's position and the speed from which the braking trajectory is predicted.
When CAM is received from a tram in the surroundings, OBU analyzes, based on the digital map, if the second tram is at the same rail track (i.e., a collision can occur).
If the second tram is at the same rail track, the OBU then predicts the trajectory of the second tram by time propagation of the received state.

The reason why the OBU predicts the braking trajectory of the first tram is to warn the driver only in situations in which the braking is necessary to avoid a collision, which, in turn, reduce the number of unnecessary (or premature) warnings.

\label{sec:method_Desc}

\section{Field Tests}
In this paper, we present the first stage of field tests in which we tested the proposed RCAS in scenarios of imminent rear-end collision with a stationary leading tram.
Another experiments which will test other potential collision situations are planned for mid-year 2020.

In this simplified scenario, the prediction of collision is reduced only to the evaluation of braking distance $d_\mathrm{br}$ of following tram and distance gap $d_g$ to the leading tram.
More precisely, the collision warning is triggered if:
\begin{equation}\label{eq:collsion_trigger_cond}
d_\mathrm{br} + d_s + v_kt_r \geq d_g\;,
\end{equation}
where $d_s$ is a safety margin, $t_r$ is a reaction time of a driver and $v_k$ is currently estimated speed of the tram.
We tested the proposed RCAS system in tram depot with real following tram and leading (stationary) tram simulated by a car (equipped with V2X communication module) parked next to the track.
The rear-end of simulated tram is for validation purposes marked by cones, see Fig.~\ref{fig:side_view_tram_stop_30kmh}.

\subsection{Initialization}
For experiments, we used a VarioLF tram developed by a company Pragoimex.
We list some relevant parameters regarding the tram and the algorithm in Tab.~\ref{tab:varioLF_param}.
The used reaction time $t_r$ corresponds to a focused driver.  
Also, we set zero safety margin $d_s$ to clearly evaluate how precise the braking distance prediction is. 
In a real deployment, $d_s$ would be of course positive.
\begin{table}[b]
\caption{Relevatnt parameters of the VarioLF tram and the algorithm.}\label{tab:varioLF_param}
\begin{center}
\begin{tabular}{ccc}
Parameter 				& Notation			& Value 					\\\hline
Curb weight				& -					& $\SI{21200}{\kilo\gram} \pm \SI{5}{\percent}$  	\\
Wheel mass				& $m_w$				& $\SI{195}{\kilo\gram}$		\\
Wheel radius 			& $r$				& $\SI{350}{\milli\meter}$ 	\\
Maximum speed 			& - 				& $\SI{65}{\kilo\meter\per\hour}$ \\
Total power of motors	& $P_\mathrm{max}$	& $4\times \SI{90}{\kilo\watt}$		\\
\hline
Reaction time			& $t_r$				& $\SI{1}{\second}$					\\
Safety margin			& $d_s$				& $\SI{0}{\meter}$					\\
Error covariance		& $P_0$				& $I_{3\times3}$	\\
Model covariance		& $q$				& $1$					\\
Measurement covariance  & $[\sigma_1, \sigma_2, \sigma_3]$	&	$[25,0.25,0.1]$	\\
\hline
\end{tabular}
\end{center}
\end{table}

\begin{figure}[tb]
	\centering
	\includegraphics[width=8.4cm]{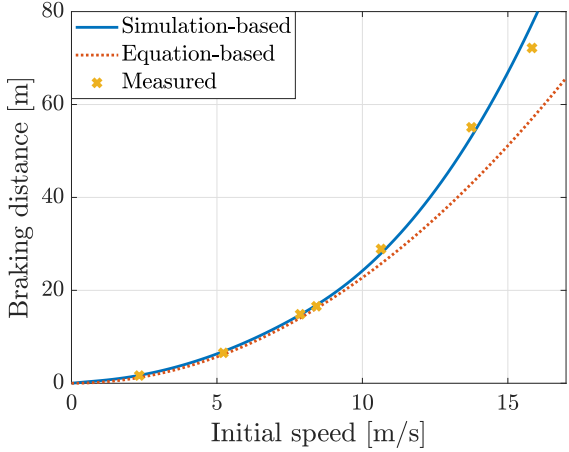}
	\caption{Comparison of measured braking distance with simulation-based braking distance prediction and equation-based computation.}\label{fig:braking_dist_comparison}
\end{figure}

With the given tram, we first conducted experiments consisting of maximal service braking from different initial speeds to identify parameters of the model described in Sec.~\ref{sec:braking_traj_pred} for an accurate prediction of the braking distance.
Identification was done by setting unknown parameters (weight $M$, traction constant $K_t$, and adhesion parameters) of the model to match the (simulated) deceleration with the measured deceleration during braking.
With a partial knowledge of parameters of the tram and condition of rail track, we identified the parameters of the model as:
\begin{itemize}
\item $M=\SI{21200}{\kilo\gram}$ (the same as curb weight of VarioLF tram since only three people were in the tram during experiments).
\item $a_a=0.54$, $b_a=1.2$ and $c_a=d_a=0.2$ (adhesion parameters for slightly wet conditions~\cite{takaoka_disturbance_2000} caused by a moderate rain few hours before the experiments).
\item $K_t = \num{2352}$.
\end{itemize}
We assume that adhesion parameters remained constant during all experiments.
Also, the rail track in the depot is almost tangent (with negligible slope), therefore $\theta = 0$ at every point on the rail track.
See Fig.~\ref{fig:braking_dist_comparison} for comparison of measured braking distances (from all experiments) with the braking curve generated from the simulation with identified parameters.
We also display, only for comparison, a braking curve computed by an equation $d_\mathrm{br}= 0.5v^2|a_\mathrm{br}|^{-1}$ with a maximal deceleration $|a_\mathrm{br}| = \SI{2.2}{\meter\per\second\squared}$.

\subsection{Results}
\begin{figure}[tb]
	\centering
	\includegraphics[width=8.4cm]{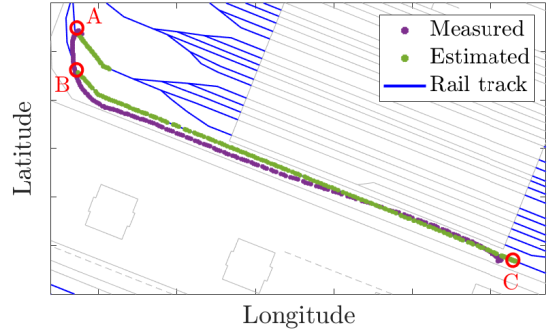}
	\caption{Coparison of position measured by the GNSS module and output of the estimation. Points A and C depicts starting end final positions, respectively.}\label{fig:estimation_comparison_position}
\end{figure}

We tested the RCAS with the following tram approaching the stationary tram consecutively at different speeds ranging from $\num{10}$ to $\SI{60}{\kilo\meter\per\hour}$.
First, we focus separately on the estimation of the position and the speed for the experiment at $\SI{50}{\kilo\meter\per\hour}$ to show several properties of the estimation.
The comparison of estimated and measured positions is shown in Fig.~\ref{fig:estimation_comparison_position}.
We can see that the use of the digital map is essential for correct localization on track.
However, at track switches, the estimation algorithm had chosen the wrong branch of the track (as between points A and B in Fig.\ref{fig:estimation_comparison_position}), which resulted in incorrect localization.
This could lead to some situations in the wrong prediction of an imminent collision.

\begin{figure}[tb]
	\centering
	\includegraphics[width=8.4cm]{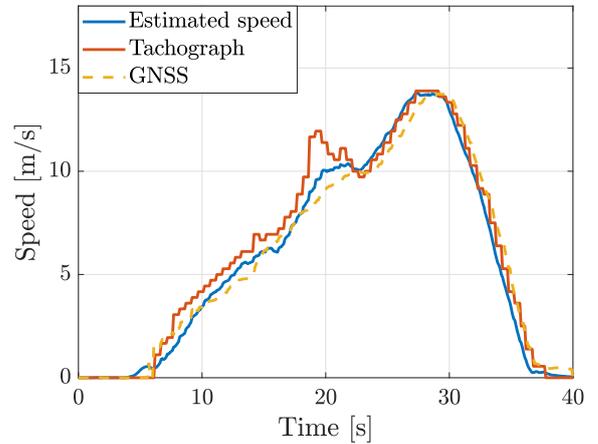}
	\caption{Comparison of estimated speed with sensor measurements for the experiment at $\SI{50}{\kilo\meter\per\hour}$.}\label{fig:estimation_comparison_speed}
\end{figure}

The comparison of estimated and measured speed (from the GNSS module and tachograph) is shown in Fig.~\ref{fig:estimation_comparison_speed}. 
As can be seen from the plot of tachograph speed at time $\SI{18}{\second}$, the tram experienced a wheel slip when accelerating (the speed from tachograph was too high).
On the contrary, the estimated speed followed the GPS speed, so it remained close to the true speed. 
This confirms that our algorithm can deal with the wheel slip, which occurs quite commonly.

\begin{figure*}[tb]
	\centering
	\includegraphics[width=17.7cm]{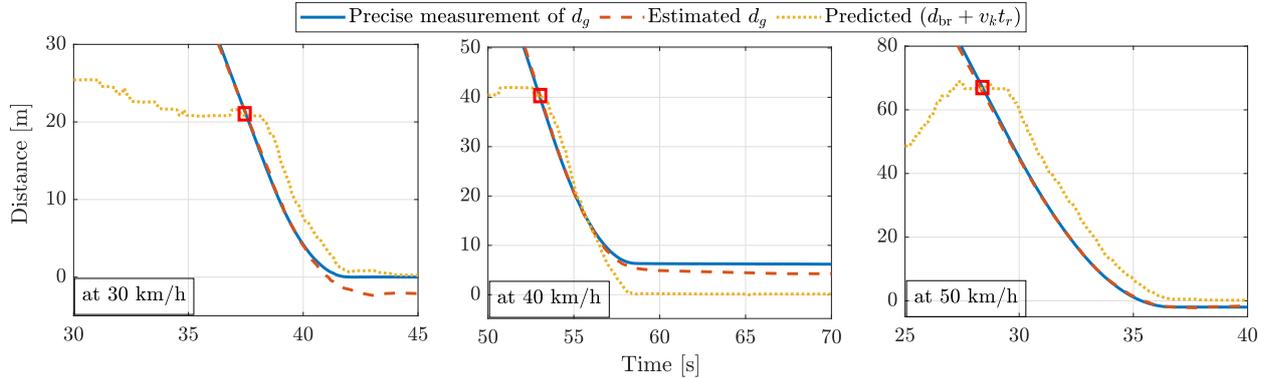}
	\caption{Results from three experiments at three approaching speeds. Red squares depict the moment of collision warning where estimated distance gap $d_g$ was lower than sum of braking distance $d_\mathrm{br}$ and reaction distance $v_kt_r$ at current speed $v_k$ with reaction time $t_r$. We also compare estimated distance gap with precisely measured (using \textit{u-blox ZED-F9P} high precision GNSS module) distance gap.}\label{fig:collision_prediction_30_40_50kmh}
\end{figure*}

\begin{table}[t]
\caption{Analysis of RCAS: paricular values for each experiment}\label{tab:values_experiment}
\begin{center}
\begin{tabular}{cccc}
$v_\mathrm{appr}~[\si{\kilo\meter\per\hour}]$	& ${d}_w~[\si{\meter}]$			& $\bar{t}_r~[\si{\second}]$ & $\bar{d}_g~[\si{\meter}]$	\\
\hline
$28.8$											& $ 20.8 $						& $0.7$						 & $0$	\\
$38.9$											& $ 40.4 $						& $0.5$  					 & $6.2$	\\
$49.8$	 										& $ 60.7 $						& $1.2$						 & $-1.6$	\\
\hline
\end{tabular}
\end{center}
\end{table}

The analysis of the collision avoidance system is given for experiments at approximately \numlist{30;40;50}$\,\si{\kilo\meter\per\hour}$.
During each experiment, the tram reached the desired approaching speed $v_\mathrm{appr}$.
Collision warning (interfaced as a high-pitch sound signal) was then given at the estimated distance gap $d_w$.
The real reaction time of the driver was $\bar{t}_r$.
The tram then stopped with precisely measured distance gap $\bar{d}_g$ (negative value means overshoot).
Particular values for each experiment are given in Tab.~\ref{tab:values_experiment} and plot of estimated quantities of condition~\eqref{eq:collsion_trigger_cond} are in Fig.~\ref{fig:collision_prediction_30_40_50kmh} with depicted moment of collision warning.
The record from dash-cam during experiments is available at~\href{https://youtu.be/OwIty4vUYT8}{https://youtu.be/OwIty4vUYT8}.

\begin{figure}[t]
	\centering
	\includegraphics[width=8.4cm]{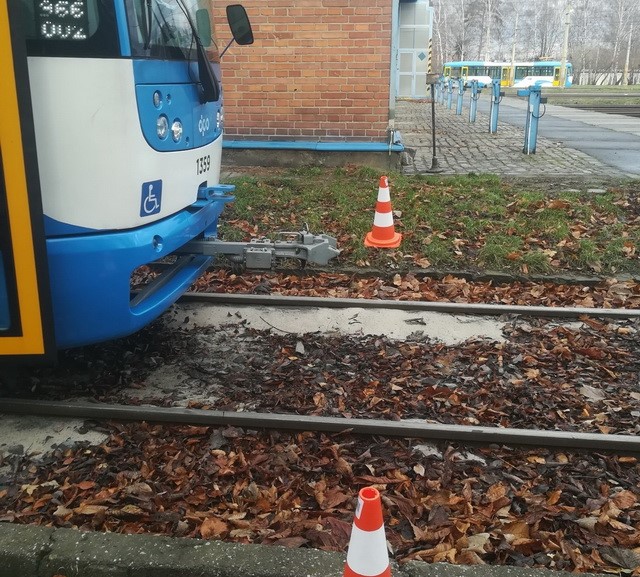}
	\caption{Side view of the tram stop position from the experiment at $\SI{30}{\kilo\meter\per\second}$. Cones represent position of rear-end of simulated stationary (leading) tram.}\label{fig:side_view_tram_stop_30kmh}
\end{figure}

The best result was at $\SI{30}{\kilo\meter\per\hour}$ where $\bar{d}_g = 0$, see Fig.~\ref{fig:side_view_tram_stop_30kmh}.
At $\SI{40}{\kilo\meter\per\hour}$, the real reaction time of the driver was lower than set reaction time, which results in a stop with undershoot whereas at $\SI{50}{\kilo\meter\per\hour}$ a bit higher reaction time results in an overshoot.
In the real deployment of the RCAS, safety margin $d_s$ must be thus set to warn the driver beforehand.

\section{Conclusion and Future Development}
In this paper, we described the design, implementation, and testing of a rail collision avoidance/warning system for trams. The system is based on vehicle-to-vehicle (V2V) communication of the motion states (positions, velocities) and their predictions among vehicles. The algorithm was implemented in the onboard vehicle-to-everything (V2X) communication units, which are already installed in every vehicle of the public transportation operator fleet.
We have successfully tested the system in several test runs, and we observed that the driver was warned at the right moment to start full braking. As an easy-to-implement extension of the already existing V2X infrastructure, our approach turns out a way to increase the safety of public transportation with low cost and no additional hardware.  

The fact that the whole tram fleet in Ostrava is equipped with V2X onboard units allows a large-scale deployment of the collision avoidance system (planned for the second half of 2020). 
As the algorithm in the experiment was already implemented in the standard OBU, a transition from this experimental verification to a full-scale deployment should be a straightforward task. 
One of the next steps would, therefore, be an assessment of the reliability of the proposed collision avoidance/warning system. A methodology for such evaluation should have to be developed.

A few technical issues remain to be solved to make the system fully deployable. 1) All our experiments were conducted in the depo where the track slope was negligible, and although the model accounts for the slope, the implementation must properly combine the measurements from the accelerometer and the digital map (with the track slope included). 2) A future test will have to include a nonstationary leading tram. This could also be realized using a dummy (virtual) tram. 3) Better projection of the position at track switches should be implemented. This might be achieved by using the Interacting multiple model algorithm~\cite{mazor_interacting_1998}.

The collision avoidance system could be extended to include other vehicles of Public transportation company fleet (buses, trolley-buses) or even cars (with increasing penetration of V2X technology).

A reliable prediction of a braking distance will also be a base for implementing an intersection signal violation warning when at least some intersection controllers are able to send SPAT and MAP messages about the signal plan via their road-side unit (RSU). 
Since the signal plan is available in RSUs, such tests may be straightforward.

\addtolength{\textheight}{-12cm}   

%

\section*{ACKNOWLEDGMENT}
We would like to acknowledge the help from engineers of Ostrava Public Transit Co. Inc. (Dopravní podnik Ostrava, a.s.).


\bibliographystyle{IEEEtran}
\bibliography{biblio}

\end{document}